\documentclass[useAMS,usenatbib]{mn2e}

\usepackage {graphicx}

\title[Oxygen abundance distribution in Galactic disc]{Oxygen abundance distribution in Galactic disc.}

\author[S.A. Korotin et al.]
{S.A. Korotin$^{1}$\thanks{E-mail: serkor@skyline.od.ua}, 
S.M. Andrievsky$^{1,2}$,
R.E. Luck$^{3}$,  
J.R.D. L\'epine$^{4}$,
\newauthor
W.J. Maciel$^{4}$ 
and V.V. Kovtyukh$^{1}$
\\
$^1$Department of Astronomy and Astronomical Observatory, Odessa
National University\\ and Isaac Newton Institute of Chile, Odessa Branch,
Shevchenko Park, 65014 Odessa, Ukraine\\
$^2$GEPI, Observatoire de Paris-Meudon, CNRS, Universite Paris Diderot, 92125
Meudon Cedex, France\\
$^3$Department of Astronomy, Case Western Reserve University 
10900 Euclid Avenue, Cleveland, OH 44106-7215\\
$^4$Instituto de Astronomia, Geof\'{\i}sica e Ci\^encias Atmosf\'ericas da
Universidade de S\~ao Paulo,\\ Cidade Universit\'aria, CEP: 05508-900, 
S\~ao Paulo, SP, Brazil
} 

\begin{document}

\date{Accepted. Received ; in original form }

\pagerange{\pageref{firstpage}--\pageref{lastpage}} \pubyear{2014}

\maketitle

\label{firstpage}

\begin{abstract}
We have performed a NLTE analysis of the infrared oxygen triplet for a
large number of cepheid spectra obtained with the Hobby-Eberly
telescope. These data were combined with our previous
NLTE results for the stars observed with Max Planck Gesellschaft
telescope with the aim to investigate oxygen abundance distribution
in Galactic thin disc. We find the slope of the radial (O/H) gradient value 
to be equal --0.058 dex/kpc. Nevertheless, we found that there could be a hint
that the distribution might become flatter in the outer parts
of the disc. This is also supported by other authors who
studied open clusters, planetary nebulae and \mbox{H\,{\sc ii}} regions. 
Some mechanisms of flattening are discussed.
\end{abstract}


\section{Introduction}
From the astronomical point of view oxygen is among the most interesting 
elements in the Universe. This is because it is the third most abundant element, 
it constitutes the base of the life (at least on the Earth), and 
it is used as a proxy of the global metal content in several astronomical objects, 
thus,  when measured in objects with different ages, it is a tracer of the temporal 
evolution of the chemical content of our Galaxy. General information about the 
oxygen abundance distribution in Galactic substructures comes from spectroscopic 
study of sources such as stars, planetary nebulae, \mbox{H\,{\sc ii}} regions, and
interstellar matter. Comparison of the data on oxygen abundance provided by these 
sources shows that there are discrepancies that affect our understanding of 
those processes that are responsible for the production and distribution of this 
element in our Galaxy and other galaxies. At present, new accurate oxygen abundance 
data from a large homogeneous sample of objects at the different Galactocentric 
distances are urgently necessary. 

Since oxygen and other $\alpha$-elements are produced in explosive processes 
in SNe~II, it is of particular interest to study the distribution of this element 
in Galactic thin disc. Characteristics of such a distribution may reveal the spatial 
dependence of the SNe~II activity that, as we believe, is caused by the  
interstellar gas density distribution in the disc, the efficiency of the Galactic 
spiral arms' influence on the gas, and the local metallicity level. 

The current status of the abundance gradient studies with 
Galactic cepheids indicates growing evidence that elemental 
distributions in the disc require, for their description not a single slope gradient, 
but instead a bimodal (or even multimodal) structure with 
different slopes in each region. For instance, \citet{An02b} using a sample of 
cepheid stars demonstrated that among other $\alpha$- and iron-group elements, 
oxygen and iron show a steeper increase of their content toward the Galactic 
centre. This increase at about 6.6 kpc changes the flat distribution in the 
solar vicinity to the much steeper one in the range of about 4-7 kpc. This is 
also supported by the results of \citet{Ped09} and \citet{Gen13} (both studies 
deal with iron distribution).

Later, \citet{Lu03} reported about a clear separation in elemental abundance 
distributions of the outer Galactic disc from its middle part. This separation 
is associated with Galactocentric distance of about 10 kpc. This finding was 
confirmed by \citet{AND04} with a larger sample of stars. They 
concluded that "the wriggle feature in the metallicity distribution, which is 
associated with Galactocentric distance of about 11 kpc can be interpreted as a 
change of metallicity level in vicinity of the Galactic corotation resonance".

In contrast, \citet{Kov05}, and \citet{Lu06} did not find any strong reason 
to not use a simple linear gradient to describe the spatial abundance distributions 
obtained with even more expanded sample of the cepheid stars. The same 
conclusion was made by \citet{luc11} and \citet{luclam} (but they noted an 
increased scatter of the abundance data outward the distance of 10 kpc). In part,
this conclusion may be affected by including the new cepheids covering more
distant parts of the disc in Galactocentric longitude, that could clean up any 
specific features in the radial abundance distribution.

\citet{Gen14} have also used a single gradient value to describe the iron 
abundance distribution over a broad range of Galactocentric distances from 5 to 
19 kpc. \citet{Lem13} also do not support the flattening of the abundance 
distribution for some elements in the outer Galactic disc, but at the same time
they do not provide the corresponding data for oxygen and iron.

As we can see, the situation with characteristics of the abundance distribution 
in Galactic disc of such astrophysically important elements as oxygen and iron 
is far from being clear. All the data on oxygen distribution from the giant and 
supergiant stars F-G-K stars come from analysis of either forbidden oxygen lines 
at 630.03 and 636.38 nm, or 615.6-615.8 triplet, but  there are known problems 
of the reliable oxygen abundance determination from this restricted sample of the 
lines. For example, two forbidden lines can hardly be detected in the supergiant 
spectra if the effective temperature of the star is higher than 5500 K, but this
is the case for a large amount of cepheids. Moreover, these lines are often 
polluted with terrestrial absorptions. Since this region is dominated by 
terrestrial bands, the additional measures of the spectra cleaning are needed. 
Among the triplet lines, only the line 615.8 nm is more or less detectable in 
the supergiant spectra. It cannot be detected in the medium-to-high resolution 
spectra of the stars with effective temperature lower than 5500-6000 K. Thus, 
in some cases, in order  to derive oxygen abundance for a large sample of the 
cepheid stars one needs to use  either forbidden lines (for cooler stars, or 
for phases of minimum with lower temperature), or triplet lines (for the hotter 
stars).

A good solution would be to use of the 777.1-777.4 nm triplet. It affords the 
most suitable opportunity to derive oxygen abundance in F-G-K supergiant stars 
because: 1) its lines are quite strong and very well shaped over a wide range
of the effective temperature, 2) these lines are practically not blended, 
3) they are reachable with many spectrograph even for the faintest cepheids.
The most significant obstacle, which prevents a wide use of this triplet in 
spectroscopic analyses, is an obvious necessity to apply sophisticated NLTE 
approximation in order to get correct oxygen abundance. This is a well know 
problem (see e.g. a very instructive example about extremely strong dependence 
of the derived oxygen abundance on the effective temperature  provided by the 
LTE study of \citealt{Sch06}). 

The main goal of the present paper was to derive correct NLTE oxygen abundance 
for a large homogeneous sample of cepheids using the IR oxygen triplet, and 
then to construct on this basis the oxygen abundance distribution in the 
Galactic thin disc.

The paper is organized as follows. Our sample of cepheids is presented in 
Sect. 2. The NLTE approximation that was used to derive oxygen abundance
from the strong IR triplet is detaily described in Sect. 3. The
oxygen abundance distribution in the disc is the subject of Sect. 4.
The Discussion and Conclusion summarize our results from the point of 
view of the recent observational and theoretical works on abundance 
gradients in the Galactic disc.

\section{Our sample of stars}

Continuing our program of the Galactic abundance gradient investigation 
\citep{AND02,An02b,An02c,Lu03,AND04,Lu06,Le11}, we study the oxygen 
distribution in the thin disc with a large sample of Galactic cepheids. Our 
sample consists of two subsamples. One contains the Hobby-Eberly Telescope 
(HET) data described in detail by \citet{luclam}. The second is composed by 
the MPG Telescope data, which are described in \citet{luc13}. The observational 
data for both samples were collected by one of the authors (REL). The list of 
the studied stars (only HET data) and their spectra can be found in Table 
\ref{Tabsam} (electronically available in its full form).

\begin{table*}
\begin{center}
\caption[]{Parameters and Abundances for Program cepheids (only header of the 
table is showed. The full table is available in electronic form)}
\label{Tabsam}
\begin{tabular}{cccccccc}
\hline
Object & Phase &T$_{\rm eff}$, K & $\log~g$&  V$_{\rm t}$. km~s$^{-1}$&	(Fe/H) & (O) & $\rm R_{G}$ (kpc)\\
\hline
AA Gem & 0.422 & 5141 & 1.23 & 3.92 & 7.34 & 8.51 &  11.55\\
AA Gem & 0.658 & 5190 & 1.45 & 5.85 & 7.38 & 8.51 &  11.55\\
AA Mon & 0.841 & 5797 & 2.26 & 4.90 & 7.41 & 8.51 &  11.36\\
\hline
\end{tabular}
\end{center}
\end{table*}

\section{Method of analysis} 

The NLTE approximation was used to derive oxygen abundance in our stars. For 
this we used the \mbox{O\,{\sc i}} triplet : 777.4 nm. Study of the NLTE 
deviations in the IR oxygen triplet was initiated by many authors in the past 
\citep[e.g.][] {kis91,tak92,CarJ93,Paun99,Ree99,mis00,Pr00,Fab09,Sit13}. 
As shown by those authors NLTE effects significantly 
strengthen these lines. At the same time available \mbox{O\,{\sc i}} 
atomic models were not always able to reproduce observed triplet profiles. For 
instance, \citet{Pr00} found different oxygen abundance in A and F stars derived 
from IR lines and lines in visual part of the spectrum.

The most complete atomic models published by \citet{Sit13} and \citet{Fab09}
take into account the most recent collisional rate values between
atoms and electrons \citep{barklem07}. Nevertheless, the lack of  detailed 
calculations describing collisions with hydrogen atoms force the use of Drawin's 
formula \citep{drawin} with a very uncertain correcting factor 
varying from 0 to 1.

\begin{figure}
\resizebox{\hsize}{!}
{\includegraphics{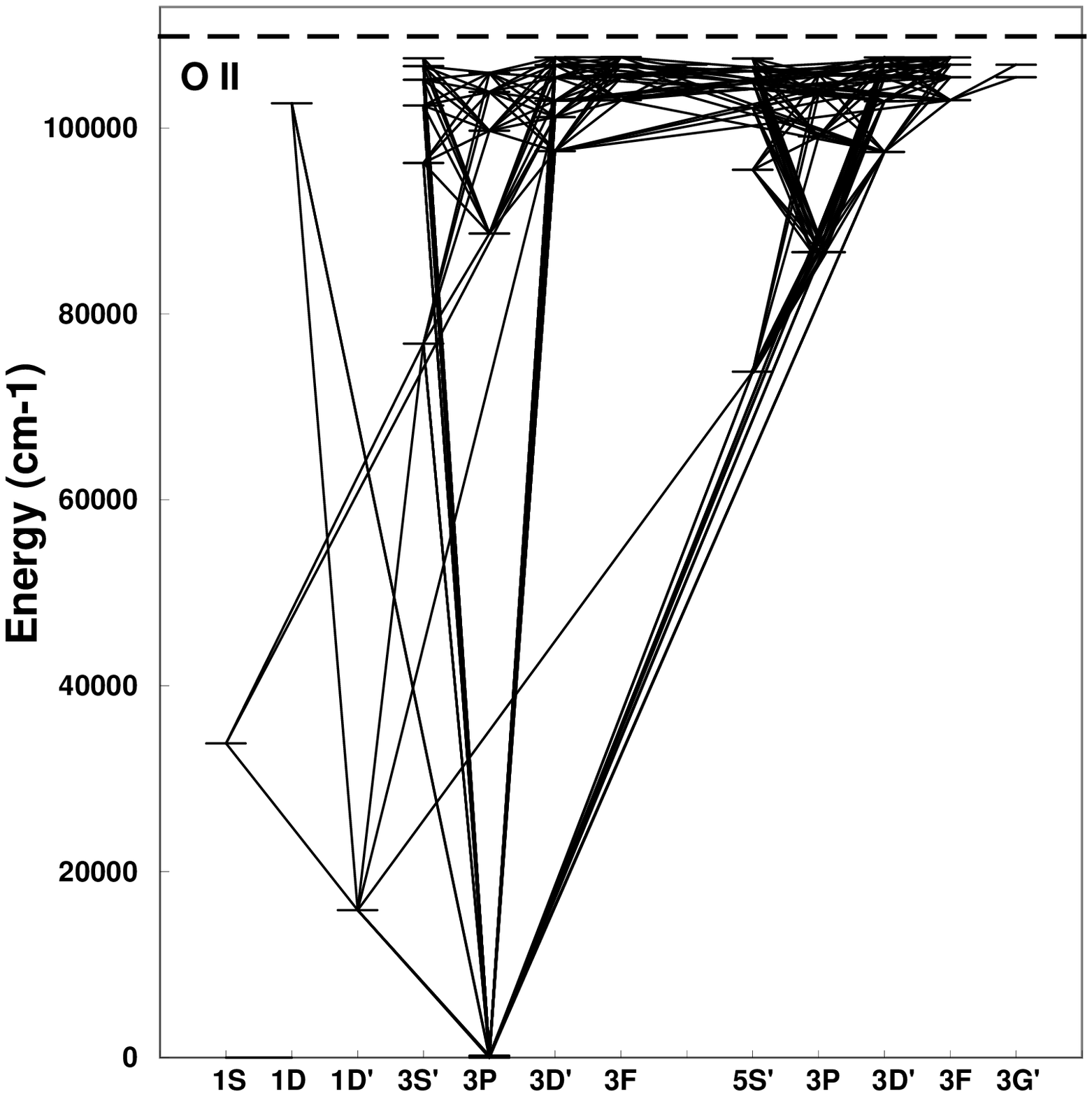}}
\caption[]{The Grotrian diagramme for \mbox{O\,{\sc i}}.}
\label{GrDiag}
\end{figure}

In this work we modified our oxygen atomic model first presented in 
\citet{mis00} and then updated in \citet{dob14}.  
Our present model consists of 51 \mbox{O\,{\sc i}} levels of singlet, triplet and 
quintet systems and the ground level of \mbox{O\,{\sc ii}} ion. An additional 24 levels 
of neutral oxygen and 15 levels of ions of higher stages were added for 
particle number conservation. The Grotrian diagram of our model is shown in
Fig. \ref{GrDiag}. Fine-splitting was taken into account only for the ground 
level and 3p5P level (the upper level for 777.4 nm triplet).

Two-hundred and forty eight bound-bound transitions were included in 
the detailed analysis. Photoionization cross-sections were selected from 
TOPBASE \citep{CMO93}. Accurate quantum-mechanical calculations were employed 
for the first 19 levels to find collisional rates with electrons 
\citep{barklem07}. Using spline interpolation the collision rates can be interpolated 
against temperature. The remaining allowed 
collisional transitions were approximated by the \citet{regemorter} formula, 
while for the forbidden transitions we applied the \citet{allen73} formula with the 
collision strength set equal to 1.0. Collisions with hydrogen atoms were 
described using Drawin's formula \citep{drawin}, as modified by 
\citet{SH84}. An oscillator strength of f=0.001 was used for the forbidden transitions 
\citep{Fab09}. Generally, our model is quite similar to that of 
\citet{Sit13}. The MULTI code \citep{carlsson} was used to compute non-LTE 
level populations. This code was modified by \citet{KAL99}. 

As shown by \citet{Sit13}, in order to achieve agreement between 
oxygen abundances in A stars derived from IR triplet lines, and from the lines of the 
visual part of the spectrum, it is necessary to decrease the collisional rates 
calculated by \citet{barklem07} by a factor of four. In order to adjust oxygen 
abundances from different multiplets in solar spectrum, and in the Vega spectrum, they 
adopted a correcting factor in the Drawin's formula $S_{H}=1$.  
We performed test calculations for Vega and Sirius using our oxygen atom model 
and found that it is not possible to describe observed line profiles for 
different multiplets with a single oxygen abundance value. While the \mbox{O\,{\sc i}} 
triplet at 615.7 nm is almost free of the NLTE influence, the lines of triplet 
at 777.4 nm appear to be weaker than the observed ones.  

If we decrease by four times the collisional rates calculated by \citet{barklem07}, as 
done by \citet{Sit13}, we can fit the observed profiles for these 
two triplets with a single abundance, but the synthetic lines of another IR 
triplet at 844.6 nm appear to be significantly increased compared to 
observations. This is seen even in the Procyon spectrum, although in its cooler 
atmosphere the collisions with hydrogen atoms begin to be more important.

For our test calculations we used the same stellar parameters and oxygen 
abundance as \citet{Sit13}. Table \ref{Tabst} lists the source of 
observed spectra and stellar parameters.

\begin{table*}
\begin{center}
\caption[]{Stellar parameters}
\label{Tabst}
\begin{tabular}{cccccccc}
\hline
Star    & T$_{\rm eff}$, K & $\log~g$ & [Fe/H] & V$_{\rm t}$. km~s$^{-1}$ & (O/H) & Res. & Source \\
\hline
Sun     & 5777 & 4.44 &  0.0 & 1.0 & 8.71 & 350000 & \citet{KFB84}      \\
Procyon & 6590 & 4.00 &  0.0 & 1.8 & 8.73 &  80000 & \citet{BJL03}      \\
Vega    & 9550 & 3.95 & -0.5 & 2.0 & 8.59 & 100000 & \citet{Tak07}      \\
Sirius  & 9850 & 4.30 &  0.4 & 1.8 & 8.42 &  80000 & \citet{BJL03}      \\
\hline
\end{tabular}
\end{center}
\end{table*}

We consider that a decrease of all collisional rates calculated by 
\citet{barklem07} is not appropriate, therefore we decided to change the 
collisional rates only for some transitions. Thus, we decreased by a 
factor of two the rates that correspond to the IR triplet 777.4 nm, and 
increased by four times the rates that correspond to 884.6 nm triplet. We also 
strengthened the coupling between 3p3P level (the upper level for the 844.6 nm 
triplet lines) and levels of the quintet system 4s5S$^{0}$ and 3d5D$^{0}$ due 
to collisions with hydrogen atoms. For the forbidden transitions we used 
f=1.0 in Drawin's formula. With these modifications, we were able to 
adjust theoretical and observational profiles of oxygen lines of different 
multiplets in the spectra of the rather hot A-F stars, the solar-type stars, as 
well as the cooler stars. For the cooler stars we used correcting factor 
$S_{H}=1$, in agreement with \citet{Fab09}, and \citet{Sit13}.

\begin{figure}
\resizebox{\hsize}{!}           	
{\includegraphics {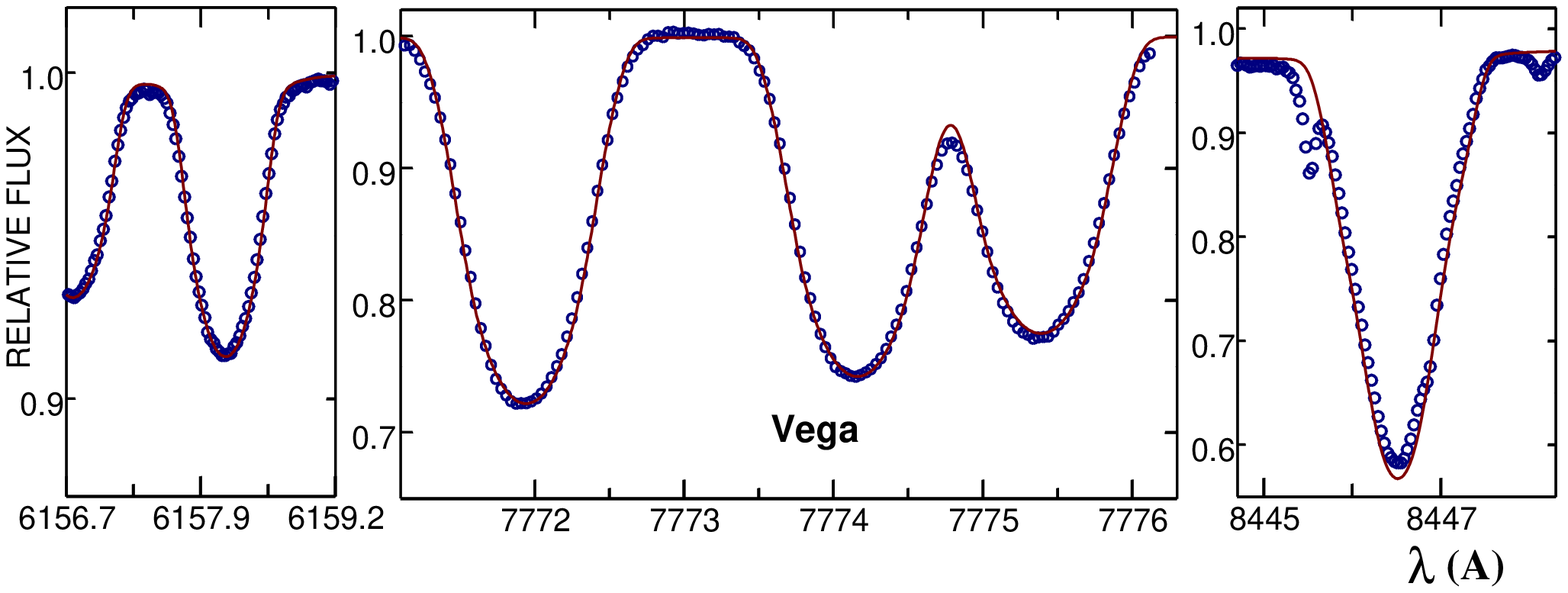}}
\resizebox{\hsize}{!}           	
{\includegraphics {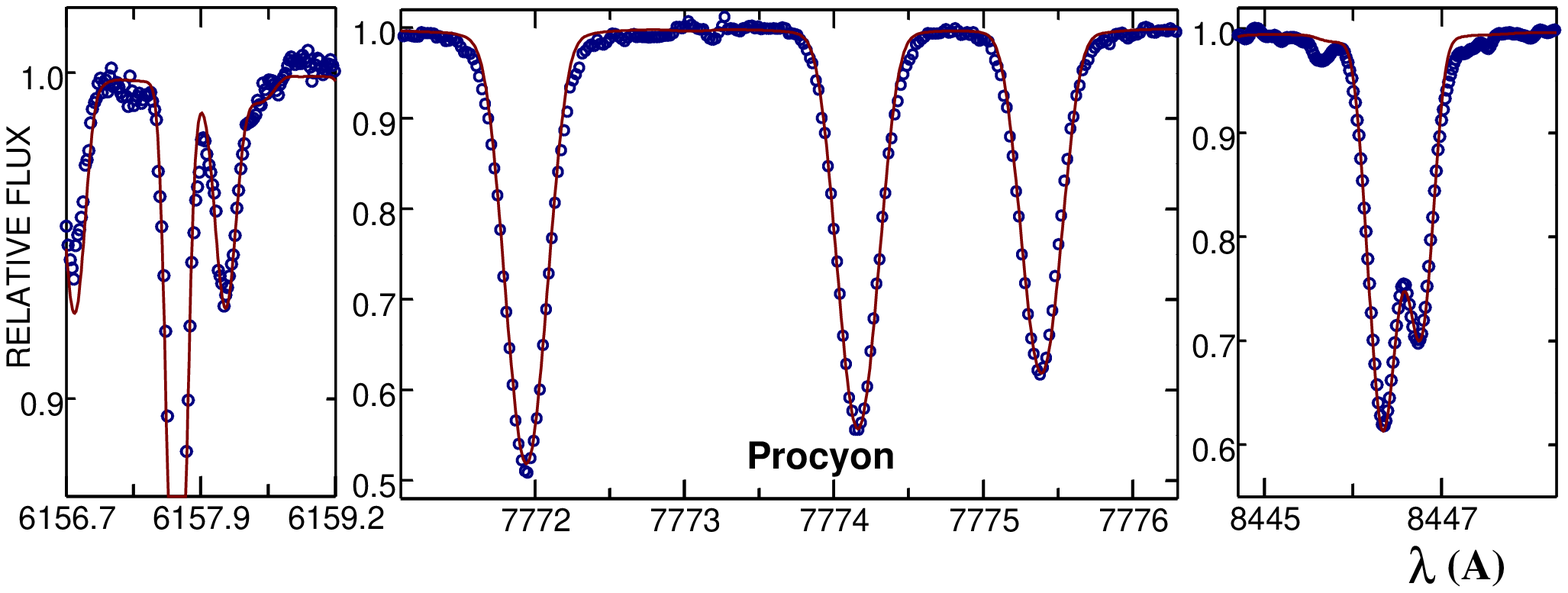}}
\caption[]{Profile fitting of the \mbox{O\,{\sc i}} line in the Vega and Procyon 
spectrum}
\label {O_star}
\end{figure}

\begin{figure}
\resizebox{\hsize}{!}           	
{\includegraphics {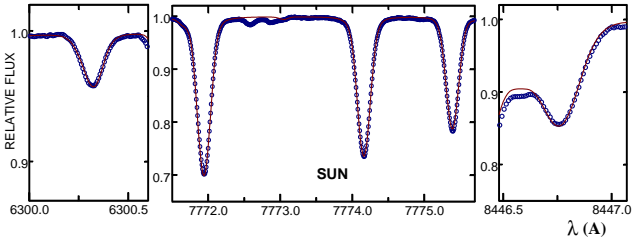}}
\caption{Profile fitting for the oxigen lines in the solar spectrum}
\label {spesol}
\end {figure}

In Fig. \ref{O_star} we show a comparison between observed and synthetic 
profiles for the lines of different multiplets for Vega and Procyon. Note that 
a single oxygen abundance value was used to synthesize all the lines for 
each star. Good agreement is seen both for 615.7 nm triplet lines that are 
not affected significantly by NLTE effects, and for IR triplets, which are 
significantly affected. Another test was applied to the solar spectrum. We 
used solar atmosphere model of \citet{CK03} that was supplemented by VAL-C 
chromosphere model \citep{VAL81} with an explicit distribution of the 
microturbulent velocity. (Note that the influence from the chromosphere on the lines 
of IR triplets is very small, no more than 1.5\% in their equivalent widths). 
Observed profiles were taken from Solar flux atlas \citep{KFB84}. We 
also made a comparison for the solar disc center using the Solar atlas of 
\citet{DRN73}. A comparison of the theoretical profiles for 630 nm 
line, the lines of 777.4 nm and 844.6 nm triplets with observed profiles in the 
solar spectrum  is shown in Fig. \ref{spesol}. To synthesize 
all the lines we used a single oxygen abundance $\log \epsilon(O) = 8.71$ (the 
same value as proposed by \citet{SAG09}).

\begin{figure}
\resizebox{\hsize}{!}           	
{\includegraphics {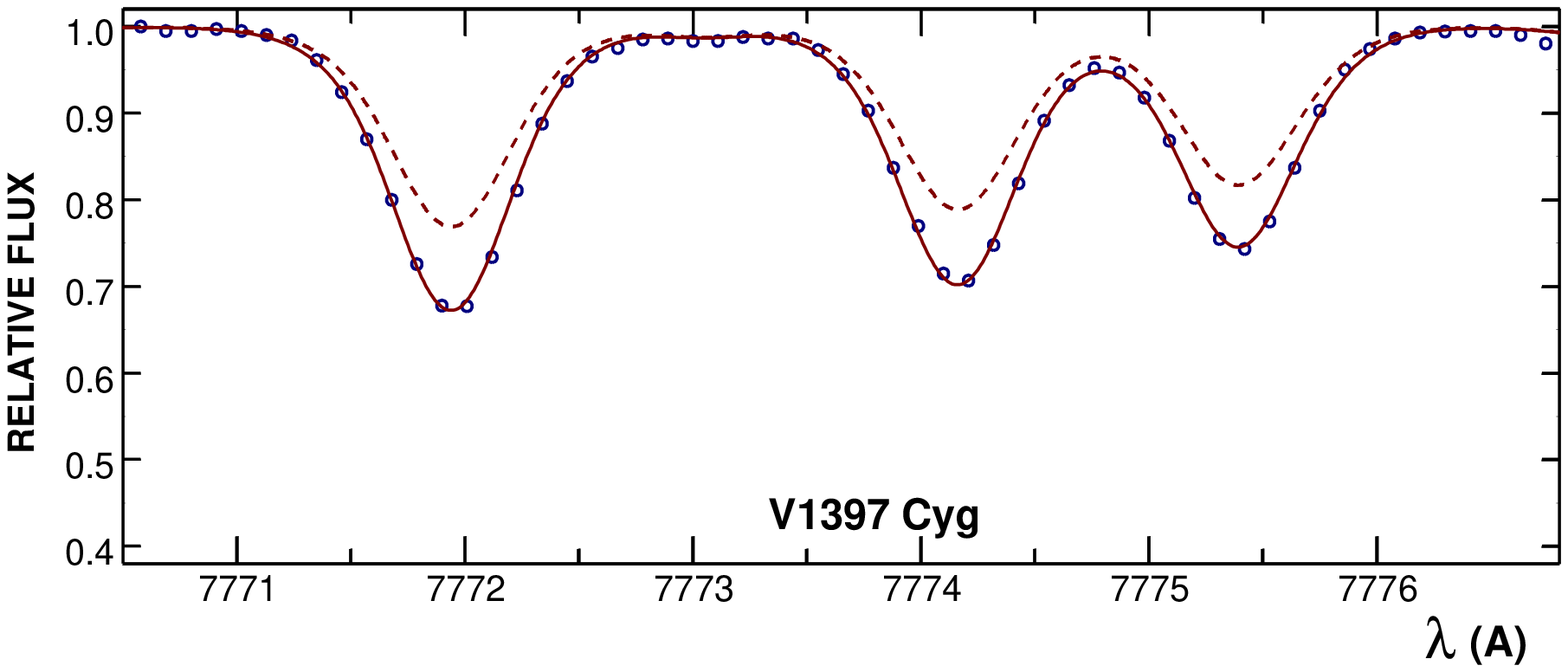}}
\resizebox{\hsize}{!}           	
{\includegraphics {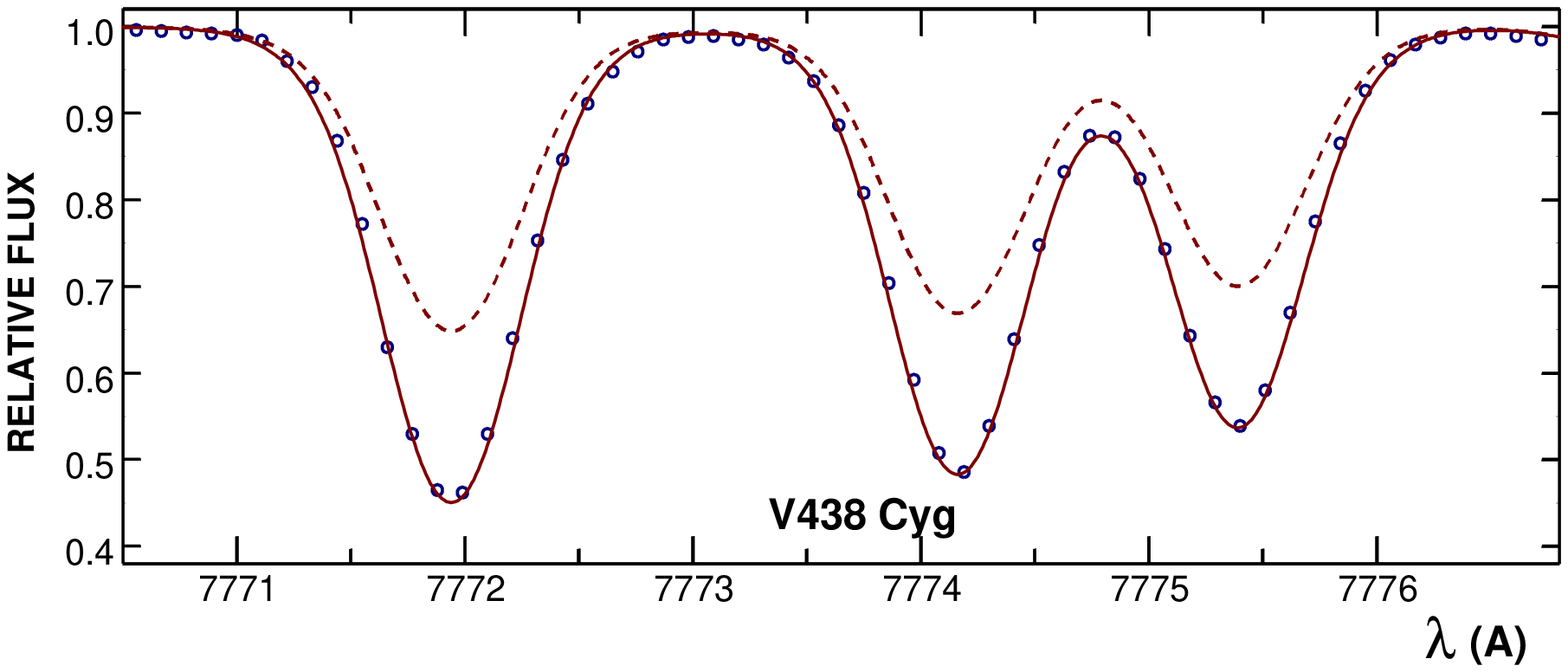}}
\caption[]{Profile fitting of the \mbox{O\,{\sc i}} line in the cepheids spectrum. $Open~circles$ -
observed profiles, $continuous~line$ - NLTE profiles, $dashed~line$ - LTE profiles calculated
with derived NLTE oxygen abundance.}
\label {Cep_O}
\end{figure}

Resulting NLTE oxygen abundances for the HET stars are gathered in 
Table \ref{Tabsam}. We also list parameters of the stars (effective 
temperature, surface gravities, microturbulent velocities) as they were derived 
in \citet{luclam} for the HET sample. Our NLTE data for the stars of MPG set can
be found in \citet{luc13}, while atmosphere parameters and LTE abundances 
for those stars were published in \citet{luc11}. Galactocentric distances 
for our program stars were calculated in the same way as in \citet{AND02}.
For the solar Galactocentric radius we used 7.9 kpc \citep{Mc00}. If we 
compare star-by-star our NLTE values with LTE values from \citet{luclam} and 
\citet{luc11}, we get the following result: NLTE-LTE = $-0.03 \pm 0.13$  
(LTE results are based on 615.6 nm line and the 630.0 nm forbidden line). 
The best fits of the NLTE synthetic profiles and observed profiles are 
given for some stars in Fig. \ref{Cep_O}. The NLTE corrections in this case 
are extremely strong, reaching sometimes 0.8-1.0 dex (the larger corrections 
are inherent to the stars with higher effective temperatures and lower gravity).

\section{Oxygen abundance distribution in the disc}

\begin{figure}
\resizebox{\hsize}{!}           	
{\includegraphics {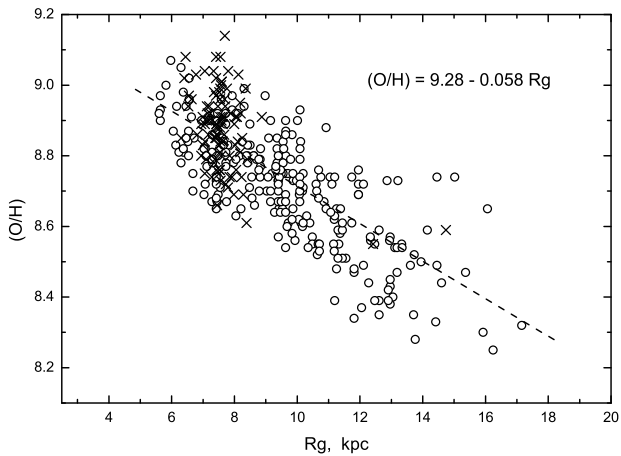}}
\caption{NLTE oxygen abundance vs. Galactocentric distance.  
$Open~circles$ - HET data, $crosses$ - MPG Telescope data.}
\label {abun}
\end {figure}

In Fig. \ref{abun} we show NLTE oxygen distribution vs. Galactocentric distance 
for the combined sample of HET and MPG Telescope stars.

The region 5-13 kpc is quite well sampled by our oxygen data and the distribution is 
rather tight. At larger distances the data are more dispersed. Samples 
observed with different telescopes produce results that agree well.

The data in Fig. \ref{abun} can be interpolated linearly. In this case we have 
the slope $\Delta (O/H)/ \Delta \rm R_{G} = -0.058$ with $R^{2} = 0.61$. 
The slope derived by Luck \& Lambert (2011) is quite similar: 
$-0.056$ with $R^{2} = 0.47$. Those authors did not discuss the possible 
bimodal character of the gradient.

\begin{figure}
\resizebox{\hsize}{!}           	
{\includegraphics {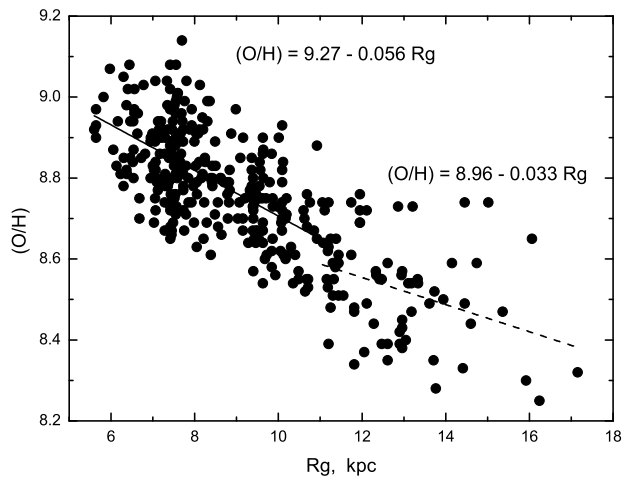}}
\caption{Bimodal distribution of oxygen abundance. 
The break position is close to 11 kpc.}
\label {bimod}
\end {figure}

Nevertheless, despite the less populated outer disc region, it seems that 
abundance distribution here could be more flat. In Fig. \ref{bimod} we show the 
same data as in Fig. \ref{abun}, but assuming a bimodal distribution. 
We formally adopted position of the possible break in the distribution at 11 kpc 
taking into account that in many studies, which report about gradient flattening, 
the distances of 10-12 kpc are mentioned (see references in Introduction and 
Discussion).

\begin{figure}
\resizebox{\hsize}{!}           	
{\includegraphics {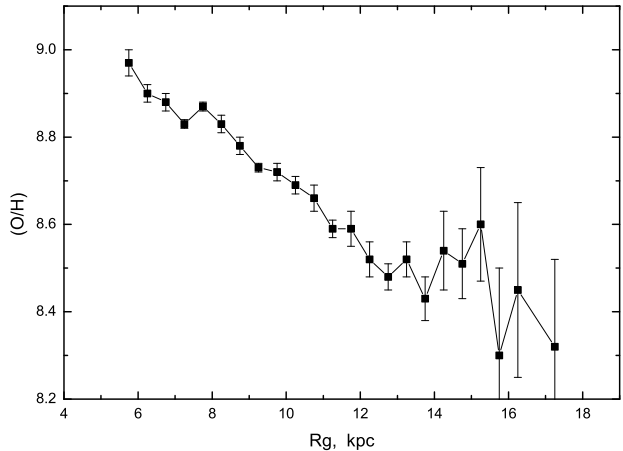}}
\caption{Oxygen abundance distribution. Binned data.}
\label {binned}
\end {figure}

More robust confirmation of this conclusion follows from Fig. \ref{binned}. In 
this figure all cepheids were binned with a step of 0.5 kpc. For each bin we 
show error of the mean. If only one star falls in a bin, then error of the mean 
was adopted to be equal 0.2. A clear change of the general trend at about 12 kpc 
and a local increase of the oxygen abundance in vicinity of 15 kpc is seen.

\section{Discussion} 

As shown by \citet{barr13}, the specific action of the corotation 
resonance results in orbital trajectory migration of stars and also in the
formation of a gas density  minimum in the corotation. For instance, their 
Fig. 7 demonstrates how a star situated initially inside the corotation 
circle after some time appears outside it. The star from outside the corotation 
circle, on contrary, migrates inside it. The typical time of this change is 
about 2 Gyr. The star migration process itself may disturb the observed radial
metallicity distribution at present time. 

More massive stars (like cepheid progenitors) have insufficient time to 
migrate far from their birth places, therefore this mechanism is not efficient 
in modifying the metallicity distribution obtained from these objects. Their 
position should be approximately associated with their parent gas clouds. 
Since the gas component flows in opposite direction from the corotation circle 
(with respect to stars) due to the influence of angular momentum loss 
and gain in vicinity of the corotation circle \citep{barr13} and references therein, 
one can expect some increase in the metallicity outwards of the corotation circle, 
which reflects a local increase in the gas density, the star formation rate, and therefore the 
metallicity production.  An element such as oxygen, being produced by 
massive short-lived stars, may show an abundance distribution that reflects 
the instantaneous mass density distribution in the Galactic disc. In our 
Fig. \ref{bimod} a peculiar region could be associated with 
$\rm R_{G} \approx 11$ kpc. \citet{barr13} find an evidence of the mass density 
gap at Galactocentric distance of about 9 kpc. This is somewhat less that 11 kpc,
but qualitatively conclusion remains the same.  

If we turn to iron abundance distribution using the most complete data for 
cepheids provided by \citet{luclam}, their Fig. 1, then we have to 
state that it is difficult to detect any peculiar point in this distribution. 
Iron is efficiently produced by SNe Ia stars, whose progenitors have a typical 
life-time of about 1 Gyr, or slightly more \citep{Ibtut}, which is 
comparable to the characteristic crossing time of particular region in vicinity 
of the corotation circle. The movement of the slightly larger number of the 
SNe~Ia progenitors from the inner part of corotation circle towards the outer 
part, and opposite movement of the smaller number of such stars from the outer 
part could increase the gas pollution with iron at the corotation circle itself 
and in vicinity of it, and therefore veil any "instantaneous" distribution 
picture of this element created by SNe of type II.

It should be noted that \citet{Daf04}, based on observations of OB stars, 
found a different behaviour of the oxygen gradient compared to the present 
work, with a strong step near the solar radius (taken as 8.5 kpc). The OB stars 
are much younger (a few million years) than the cepheids, so that in principle, 
their abundance is really that of the surrounding gas. \citet{Cos04} derived 
some flattening at Galactocenttric distances larger than 10 kpc on the basis of 
a set of Galactic planetary nebulae, and a previous work in this series 
\citep{AND04,AND13} led to the same conclusion based on cepheid variables. More
recent results by \citet{Lem13} suggest a constant slope along the Galactic 
radius, although their own data are not inconsistent with some flattening at 
large Galactocentric distances. More recent results of the same group 
based on  accurate iron abundances of a large sample of Galactic cepheids 
\citep{Gen14} are consistent with a linear gradient over a broad range of 
Galactocentric distances given by -0.06 dex/kpc, very similar to the results 
shown in Figure \ref{abun}. 

Open cluster data seem to indicate the same tendency presented in this paper, 
as recently discussed by \citet{Yon12} based on $\alpha$- and iron-group element 
abundances of new sample of objects in the outer Galactic disc. For instance, 
the plot for representative $\alpha$-element magnesiun shows a break 
of the distribution at 13 kpc with some flattening of the distribution at larger radii 
(see their Fig. 31a). In principle, similar behaviour is inherent to the cepheid 
(Mg/H) distribution (Fig. 31d). 

The same conclusion was also made earlier by \citet{mag09}, who claimed that 
there is a flattening of the iron and $\alpha$-element distribution (or even 
plateau) at the radii larger than 12 kpc. \citet{Ces07} succeeded to model 
the flattening of the gradients produced by open clusters, as well as by 
cepheids, hot stars and red giant stars.

\citet{Lep13} performing an average of abundances of several alpha elements in 
order to decrease the effect of individual errors on measurements, suggest 
(their Figure 6) that there are two distinct levels of alpha abundances, with 
some overlap in Galactic radius. 

A recent study based on Galactic \mbox{H\,{\sc ii}} regions and featuring the 
outer disc object NGC~2579 \citep{Est13} is also consistent with a flattened 
gradient at large distances from the Galactic center.

The flattening of the gradients at large Galactocentric distance seems to be an 
universal property of disc galaxies. For example \citet{Sta13} considered 
PNe and \mbox{H\,{\sc ii}} regions in the spiral galaxy NGC 300 and found that 
oxygen and other element abundance gradients from PNe are significantly 
shallower than those from \mbox{H\,{\sc ii}} regions, and this may indicate a 
steepening of the metallicity gradient in NGC 300. Very clear flattening of the 
oxygen abundance distribution in the outer parts of the spiral galaxy M83 was 
reported by \citet{bre09} (see their Fig. 9). Recently, \citet{San14} presented 
the largest and most homogeneous catalog of \mbox{H\,{\sc ii}} regions in 
several hundreds of galaxies from CALIFA survey. Many of these galaxies show a 
flattening in the oxygen abundance.

These results may in principle be affected by the time variation of the radial abundance gradients 
in the Galactic disc, but some recent work based on four different sample of Galactic planetary nebulae 
suggest that this variation was probably very small during the past 4 to 5 Gyr, so that the gradients indicated 
by these different objects are essentially indistinguishable \citep{Mac13}. Therefore, the planetary nebula 
gradient is not expected to be very different from the gradient observed in \mbox{H\,{\sc ii}} regions and 
cepheid variables, which is supported by recent additional observational data on these objects 
\citep{Fu09, Ped09}. Dynamical calculations valid for the Milky Way disc over a time span of about 
6 Gyr seem to confirm these results \citep{Cur14}. 

Additionally, some recent work on the time variation of the abundance gradients 
as a function of the redshift suggest that the variations are indeed very small 
for z $<$ 0.5 \citep[cf.][]{Gib13, Pil12}, which comprise the age bracket of most 
of the objects mentioned above.

Alternatively, the data at galactocentric distances larger than 14 kpc in 
principle may become more scattered. In such a case, the outer parts of the 
Galaxy may show abundances that reflect local events that influence the 
abundances more than coherent evolution as implied by a gradient. Unfortunately, 
this type of analysis is not really possible with cepheids (the natural lack of 
the cepheid stars at large distances). Perhaps in the future giants might be 
used to continue the project.

\section{Conclusion}

Summarizing our present study, one can note that there is a growing 
observational evidence about the flattening of $\alpha$-element distributions 
(including oxygen) in the discs of different galaxies including our Galaxy.
These data come from spectroscopic analyses of the open clusters, planetary
nebulae, H~II regions, hot stars, red giant stars, and now, from cepheids. 

Since oxygen is produced by massive short-lived stars, its abundance distribution should follow an instantaneous 
gas distribution in the Galactic disc. As recently concluded by  Barros et al. (2013) based 
on several previous theoretical studies, the radial gas density profile has a gap at the corotation circle, and 
an increased density on both sides of it.  Our observational data comprisong a large sample of cepheids, analysed taking into account NLTE effects on oxygen 
abundance determinations, seem to support this assumption, and show that 
the corotation resonance in the Galactic disc can have really significant impact on the abundance gradients in its vicinity. 
In particular,  our oxygen abundance distribution reveals a clear change of the general trend
at about 12 kpc with the subsequent flattening in the range from 12 to 15 kpc, or even small increase 
of the oxygen abundance toward 15 kpc.

\section*{Acknowledgments}
SAK and SMA acknowledges SCOPES grant No. IZ73Z0-152485 for financial support. 
Authors thank the anonymous refree for her/his valuable comments that significantly improved 
the paper.

\label{lastpage}

\bsp

\end{document}